# Ramond sector of super Liouville theory from instantons on an ALE space

Yuto Ito

*Institute of Physics, University of Tokyo,*
*Komaba, Meguro-ku, Tokyo 153-8902, Japan*

ito@hep1.c.u-tokyo.ac.jp

**Abstract**

We propose that $\mathcal{N}=2$ $U(2)$ gauge theories on the $A_1$ ALE space, with asymptotic holonomies not in $SU(2)$, correspond to the Ramond sector of super Liouville theory. As evidence, we show that the instanton partition functions for the theories with and without a fundamental hypermultiplet, computed with such holonomies, coincide with the norms or the inner products of the Whittaker vectors in the Ramond sector. This correspondence suggests that S-duality of $U(2)$ gauge theories interchanges the sectors with different boundary conditions.

# Contents

# 1 Introduction

S-duality is an equivalence of two quantum field theories. This enables us to understand strongly coupled regimes of theories from perturbative analysis of the weakly coupled theories. Some of four-dimensional $\mathcal{N} = 2$ superconformal gauge theories have S-duality and they are important examples since in these theories the instanton partition functions, which are nonperturbative objects, are exactly known and we can see how they behave under the S transformation.

In [1] it was proposed that four-dimensional $\mathcal{N} = 2$ superconformal gauge theories on $\mathbb{C}^2$ and $S^4$ with direct products of $SU(2)$ gauge symmetries are related to the Liouville field theory on the corresponding Riemann surfaces [2]. S-duality is realized as a modular transformation of the two-dimensional conformal field theory. This correspondence was extended in [3] to non-conformal $SU(2)$ gauge theories with $N_f < 4$ fundamental hypermultiplets. The instanton partition functions of these theories are shown to be equal to the norms or the inner products of the states called the Whittaker vectors in the Virasoro Verma module in Liouville theory.

More recently, in [4] this correspondence was further generalized to the correspondence between the $\mathcal{N} = 2$ $SU(2)$ pure gauge theory on $\mathbb{C}^2/\mathbb{Z}_2$ and the Neveu-Schwarz sector of $\mathcal{N} = 1$ super Liouville theory. Since there are noncontractible cycles in the asymptotic region of $\mathbb{C}^2/\mathbb{Z}_2$, the gauge field can have nontrivial holonomies $U = \exp(i \oint A)$ around them. Thus $U(2)$ gauge theories on $\mathbb{C}^2/\mathbb{Z}_2$ have four sectors, where $U = \exp(i \oint A) =$ $(1, 1), (-1, -1), (1, -1)$ and $(-1, 1)$[1]. The last two sectors do not exist in $SU(2)$ gauge theories. We will denote the instanton partition function of the sector with the holonomy $U = (e^{\pi i q_1}, e^{\pi i q_2})$ by $Z^{\text{inst}}_{q_1, q_2}$. In this case $q_1$ and $q_2$ can take values 0 and 1. In [4] it was shown that the instanton partition functions of the sectors with the holonomies $U = (1, 1)$ and $U = (-1, -1)$ are equal to the norms of the Whittaker vectors of the Neveu-Schwarz sector as follows

$$Z^{\text{pure, inst}}_{0,0} = {}_{\text{NS,even}}\langle\Lambda|\Lambda\rangle_{\text{NS,even}} \tag{1.1}$$

$$Z^{\text{pure, inst}}_{1,1} = {}_{\text{NS,odd}}\langle\Lambda|\Lambda\rangle_{\text{NS,odd}}, \tag{1.2}$$

where $|\Lambda\rangle_{\text{NS,even}}$ and $|\Lambda\rangle_{\text{NS,odd}}$ are the sums of states with integer and half-integer levels in the Verma module, respectively.

In [5] the gauge theory side of the correspondence was formulated on the $A_1$ ALE space, which is a resolution of $\mathbb{C}^2/\mathbb{Z}_2$. In [6, 7] the correspondence between the instanton partition functions of $SU(2)$ superconformal gauge theories on $\mathbb{C}^2/\mathbb{Z}_2$ or the $A_1$ ALE space and the conformal blocks in the Neveu-Schwarz sector of the super Liouville theory on the corresponding Riemann surfaces was also checked.

In [8] it was checked explicitly that the $SU(2)$ pure gauge theory on $\mathbb{C}^2/\mathbb{Z}_4$ corresponds to the parafermionic field theory with the spin 4/3-fractional supercurrent by showing that the instanton partition functions with $SU(2)$ holonomies coincide with the norms of the Whittaker vectors in the sectors analogous to the Neveu-Schwarz sector.

It is a natural question what is a counterpart of the Ramond sector in the gauge theories. We propose that the sectors with the holonomies $U = (1, -1)$ and $U = (-1, 1)$ in $U(2)$ gauge theories on $\mathbb{C}^2/\mathbb{Z}_2$ correspond to the Ramond sector in super Liouville theory.[2] As strong evidence in favor of this proposal, we will check the following relations

$$Z^{\text{pure, inst}}_{0,1} = Z^{\text{pure, inst}}_{1,0} = {}_{\text{R+}}\langle\Lambda^2|\Lambda^2\rangle_{\text{R+}} = (-i) \times {}_{\text{R-}}\langle\Lambda^2|\Lambda^2\rangle_{\text{R-}}, \tag{1.3}$$

in the case of the $U(2)$ pure gauge theory. The Whittaker vectors in the Ramond sector $|\Lambda^2\rangle_{\text{R}\pm}$ are defined in section 3. In addition, we will check the following relations

$$Z^{N_f=1, \text{inst}}_{0,1;1} = Z^{N_f=1, \text{inst}}_{1,0;0} = {}_{\text{R+}}\langle\Lambda^2/2|\Lambda, m\rangle^{(1)}_{\text{R+}} = (-i) \times {}_{\text{R-}}\langle\Lambda^2/2|\Lambda, m\rangle^{(1)}_{\text{R-}} \tag{1.4}$$

$$Z^{N_f=1, \text{inst}}_{0,1;0} = Z^{N_f=1, \text{inst}}_{1,0;1} = {}_{\text{R+}}\langle\Lambda^2/2|\Lambda, m\rangle^{(2)}_{\text{R+}} = (-i) \times {}_{\text{R-}}\langle\Lambda^2/2|\Lambda, m\rangle^{(2)}_{\text{R-}}, \tag{1.5}$$

in the case of $U(2)$ gauge theory with a fundamental hypermultiplet. The states $|\Lambda, m\rangle^{(1)}_{\text{R}\pm}$ and $|\Lambda, m\rangle^{(2)}_{\text{R}\pm}$ are also defined in section 3 and $r$ in $Z^{N_f=1, \text{inst}}_{q_1, q_2; r}$ denotes a flavor Wilson line.

---

[1] Holonomy $U = \text{diag}(e^{\pi i q_1}, e^{\pi i q_2})$ is denoted just by $U = (e^{\pi i q_1}, e^{\pi i q_2})$ in this paper.

[2] In the pure gauge theory the $U(1)$ part of $U = \mp i(i, -i)$ is decoupled since the fields are in the adjoint representation. In the theory with a fundamental hypermultiplet, the $U(1)$ part may be thought of as the Wilson line for the flavor symmetry. In either case, the total background holonomy acting on the fields gives a well-defined representation of $\pi_1(S^3/\mathbb{Z}_2) = \mathbb{Z}_2$.

As we will explain in section 2, the instanton partition functions are computed for theories with not only the holonomies of the gauge field but also the flavor Wilson line around the cycle at infinity.

As a natural extension of [1], we expect that S-duality of $\mathcal{N} = 2$ $U(2)$ superconformal gauge theories on $S^4/\mathbb{Z}_2$ corresponds to modular invariance of the super Liouville theory on the corresponding Riemann surfaces. For example, if we transform the coupling constant $\tau$ in the $\mathcal{N} = 2^*$ $U(2)$ gauge theory into $-1/\tau$, the complex structure $\tau$ of the one-punctured torus is transformed into $-1/\tau$ on the super Liouville theory side. This modular transformation exchanges the two cycles in the torus and therefore mixes the Neveu-Schwarz and the Ramond sectors. Thus the correspondence between each gauge theory sector and each super Liouville theory sector suggests that in $U(2)$ gauge theories, the S transformation $\tau \to -1/\tau$ closes only when we consider not only the sectors with the holonomies $U = (1, 1)$ and $U = (-1, -1)$ but also the sectors with the non-$SU(2)$ holonomies $U = (1, -1)$ and $U = (-1, 1)$ .

This paper is organized as follows. In section 2 we calculate the instanton partition functions of the $\mathcal{N} = 2$ $U(2)$ gauge theories with and without a fundamental hypermultiplet on $\mathbb{C}^2/\mathbb{Z}_2$. In section 3 we define the Whittaker vectors in the Ramond sector and show that the norms or the inner products of the vectors equal the instanton partition functions.

## 2 Gauge theory on the ALE spaces

In this section, we calculate the instanton partition functions of the sectors with the holonomies $U = (1, -1)$ and $U = (-1, 1)$ in the case of $\mathcal{N} = 2$ $U(2)$ gauge theories with and without a fundamental hypermultiplet on $\mathbb{C}^2/\mathbb{Z}_2$ . Before we perform this calculation in section 2.2, we review how to calculate instantons in a general $\mathcal{N} = 2$ $U(N)$ gauge theory on $\mathbb{C}^2/\mathbb{Z}_p$ following [9] in section 2.1 .

### 2.1 Instanton counting in the general case

Now let us consider an $\mathcal{N} = 2$ $U(N)$ gauge theory on $\mathbb{C}^2/\mathbb{Z}_p$, where $\mathbb{Z}_p$ acts on $(z_1, z_2) \in \mathbb{C}^2$ as $(z_1, z_2) \to (e^{2\pi i/p} z_1, e^{-2\pi i/p} z_2)$. Since there are noncontractible cycles in the asymptotic region of $\mathbb{C}^2/\mathbb{Z}_p$ , the gauge field can have a nontrivial holonomy $U = \exp(i \oint A) = (e^{2\pi i q_1/p}, \cdots, e^{2\pi i q_N/p})$ ,where $0 \leq q_1, \cdots, q_N \leq p-1$, around them. If it has the configuration with the holonomy at a saddle point in the path integral and all the fields are covariantly constant with respect to the gauge field with this configuration, then the periodicities of the fields are twisted by the holonomy. That is, running once around the noncontractible cycle on $\mathbb{C}^2/\mathbb{Z}_p$, the fields receive the gauge transformation $U$. Since the $\mathbb{Z}_p$ action rotates the fields once around the noncontractible cycle, this action causes not only the transformation $(z_1, z_2) \to (e^{2\pi i/p} z_1, e^{-2\pi i/p} z_2)$ but also the gauge transformation $U$. In order to get the partition functions of the $U(N)$ gauge theory on $\mathbb{C}^2/\mathbb{Z}_p$, we must perform path integral over the space of the field configurations on $\mathbb{C}^2$ that are invariant under this $\mathbb{Z}_p$ action.

Now we recall how to calculate the instanton partition function of the $\mathcal{N} = 2$ $U(N)$ gauge theory on $\mathbb{C}^2$. Later we will extract the contribution of $\mathbb{Z}_p$-invariant field configurations from this partition function. To regularize the volume of the moduli space of $k$ instantons, we perform a group action $U(1)^2 \times U(1)^N$ on the moduli space and consider only the contributions from the instantons corresponding to the fixed points under this group action. Choosing an element of the group $U(1)^2 \times U(1)^N$ whose generator is labeled by $\xi = (\epsilon_1, \epsilon_2, a_1, \cdots, a_N)$, we can write the instanton partition function as

$$Z(\xi) = \sum_{\text{fixed points}} \frac{1}{\prod_i w_i(\xi)|_{\text{fixed points}}} . \tag{2.1}$$

The right hand side is the sum over the fixed points and each summand is the contributions from the fluctuations around the instanton configuration corresponding to the fixed point in the moduli space. We denote the set of the weights of the action labeled by $\xi$ on the tangent space of the moduli space by $\{w_i(\xi)\}$. The equivariant parameters $a_1, \cdots, a_N$ are related to the vacuum expectation value of the adjoint scalar field in the vector multiplet as $\langle \phi \rangle = \text{diag}(a_1, \cdots, a_N)$.

The moduli space can be constructed from the two complex vector spaces, $V = \mathbb{C}^k$ and $W = \mathbb{C}^N$, and so can the tangent space at each point. Each fixed point can be distinguished as follows by the weights of the group action on the space $V$ and these weights can be described by a $N$-tuple of Young tableaux $(Y_1, \cdots, Y_N)$ where the number of the boxes is $k$, that is $|Y_1| + \cdots |Y_N| = k$. We label the box in the $i$-th column and the $j$-the row in each Young tableau by $(i, j)$. Each box corresponds to a one-dimensional subspace of $V$ and the weight of the group action on the subspace corresponding to the box $(i, j)$ in $Y_m$ is $a_m + (i - 1)\epsilon_1 + (j - 1)\epsilon_2$, where $i, j \geq 1$. The explicit formulae of the instanton partition functions of $U(N)$ gauge theories are written in appendix A.

Then let us consider the instanton partition function on $\mathbb{C}^2/\mathbb{Z}_p$. We have to perform two modifications to that on $\mathbb{C}^2$ (2.1). First we have to choose the fixed points that are invariant under the $\mathbb{Z}_p$ action and sum over only invariant fixed points. Secondly we have to choose the contributions from the fluctuations whose configurations are invariant under the $\mathbb{Z}_p$ action.

We explain which fixed point is invariant under the $\mathbb{Z}_p$ action. As explained above, each fixed point is labeled by a $N$-tuple of Young tableaux and then we will assign a charge of $\mathbb{Z}_p$ to each box in the Young tableaux. To do it, we determine a charge of each parameter of $\epsilon_1, \epsilon_2, a_1, \cdots$ and $a_N$. The parameters $\epsilon_1$ and $\epsilon_2$ are related to two subspaces $\mathbb{C}$ of $\mathbb{C}^2$ respectively and they transform under the $\mathbb{Z}_p$ action as follows

$$\epsilon_1 \to \epsilon_1 + \frac{2\pi}{p} \tag{2.2}$$
$$\epsilon_2 \to \epsilon_2 - \frac{2\pi}{p}.$$

Thus we think of the charge of $\epsilon_1$ and $\epsilon_2$ as 1 and $-1$ respectively. On the other hand, since $\{a_m\}$ $(1 \leq m \leq N)$ is the set of the weights on the space $W$, which is a fundamental representation space of $U(N)$, they transform as follows under the gauge transformation $U = (e^{\pi i q_1/p}, \cdots, e^{\pi i q_N/p})$ caused by the $\mathbb{Z}_p$ action:

$$a_m \to a_m + \frac{2\pi q_m}{p}. \tag{2.3}$$

Thus the charge of $a_m$ is $q_m$. As explained above, the box $(i, j)$ in the Young tableau $Y_m$ corresponds to a one-dimensional subspace the weight on which is $a_m + (i-1)\epsilon_1 + (j-1)\epsilon_2$. Therefore the charge of this box is $q_m + (i-1) - (j-1)$.

For example, let us think of a $U(2)$ gauge theory on $\mathbb{C}^2/\mathbb{Z}_2$ and the 2-tuple of the Young tableaux $(Y_1, Y_2) = (\{2,1\}, \{1\})$ in the following two cases. In the case of the holonomy $U = (1, 1)$, we determine the charge of each box in these Young tableaux as in $(a)$ in figure 1 and in the case of the holonomy $U = (1, -1)$, the charge of each box is as in $(b)$. For the holonomy $U = (e^{i\pi q_1}, e^{i\pi q_2})$, the box at the bottom left $(i, j) = (1, 1)$ in each of $Y_1$ and $Y_2$ has charge $q_1$ and charge $q_2$. As described in figure 1, where the charges are shown modulo $p = 2$, every time we move to the right by one step, the charge increases by one and every time we move up by one step, the charge decreases by one.

$(a)$ $(Y_1, Y_2) = (\{2, 1\}, \{1\})$    $(b)$ $(Y_1, Y_2) = (\{2, 1\}, \{1\})$
$U = (1, 1)$                    $U = (1, -1)$

Figure 1 : Examples of charge assignment to boxes in the Young tableaux

We denote the number of the boxes with charge 0 and 1 by $k_0$ and $k_1$ respectively. In the case $(a)$, $(k_0, k_1) = (2, 2)$ and in the case $(b)$, $(k_0, k_1) = (1, 3)$.

Returning to the general $U(N)$ gauge theory on $\mathbb{C}^2/\mathbb{Z}_p$, we can define $\{k_q\}$ $(0 \leq q \leq p-1)$ similarly. There is a relation between the holonomy and $\{k_q\}$ [9, 10]. Let us look at the $N$ diagonal elements of the $N \times N$ matrix $U$ representing the holonomy. We denote the number of the diagonal elements with the value $\exp(2\pi i q/p)$ by $N_q$ $(0 \leq q \leq p-1)$. The relation is as follows

$$c_1(E) = \sum_{q=0}^{p-1} \left(N_q - 2k_q + k_{q+1} + k_{q-1}\right) c_1(T_q), \qquad (2.4)$$

where $c_1(E)$ is the first Chern class of the gauge bundle $E$ and $c_1(T_q)$ is the first Chern class of a vector bundle $T_q$ whose base space is $\mathbb{C}^2/\mathbb{Z}_p$ and whose fiber space is a complex one-dimensional space. This fiber bundle keeps in account the fact that parallel transporting a section of the bundle along a noncontractible cycle on the base space gives the holonomy $e^{\pi i q/p}$. The vector bundle $T_0$ has a trivial connection and therefore $c_1(T_0) = 0$. In this paper, we only consider the case where $N = p = 2$ and $c_1(E)/c_1(T_1) = 0, 1$. In the case $c_1(E)/c_1(T_1)=0$, the relation (2.4) becomes $0 = N_1 - 2(k_1 - k_0)$ with $N_1 = 0, 2$ and in the case $c_1(E)/c_1(T_1)=1$, the relation (2.4) becomes $1 = N_1 - 2(k_1 - k_0)$ with $N_1 = 1$. That is, we consider the following three cases in this paper.

$$\begin{array}{lll} \text{Case 1} & N_1 = 0, & k_1 - k_0 = 0 \\ \text{Case 2} & N_1 = 2, & k_1 - k_0 = 1 \\ \text{Case 3} & N_1 = 1, & k_1 - k_0 = 0 \end{array} \qquad (2.5)$$

Therefore we must choose the appropriate pairs of Young tableaux when we calculate the instanton partition functions on $\mathbb{C}^2/\mathbb{Z}_2$ with the specified holonomy. For example, for the holonomy $U = (1, 1)$, or equivalently $N_1 = 0$, we must choose the pairs of the Young tableaux where $k_1 - k_0 = 0$. The pair of the Young tableaux $(Y_1, Y_2) = (\{2, 1\}, \{1\})$ in figure 1 satisfies $k_1 - k_0 = 0$ in this case. On the other hand, for the holonomy $U = (1, -1)$, or equivalently $N_1 = 1$, we must choose the pairs of the Young tableaux where $k_1 - k_0 = 0$ again. The pair of the Young tableaux in figure 1 does not satisfy $k_1 - k_0 = 0$ in this case. Therefore this pair of the Young tableaux contributes to $Z_{0,0}^{\text{inst}}$ in (1.1) but does not contribute to $Z_{0,1}^{\text{inst}}$ in (1.3).

In addition to the restriction on $N$-tuples of Young tableaux, we must restrict the weights in each summand of (2.1). For example, in the case of the $U(2)$ pure gauge theory on $\mathbb{C}^2$, the summand corresponding to the fixed point labeled by the pair of the Young tableaux $(Y_1, Y_2) = (\{1, 1\}, \{0\})$ is the product of the following eight weights

$$Z_{\{1,1\}\{0\}}^{U(2) \text{ pure}} = \frac{1}{(-a_1 + a_2)(-a_1 + a_2 - \epsilon_1)(\epsilon_1)(2\epsilon_1)(\epsilon_2)(-\epsilon_1 + \epsilon_2)(a_1 - a_2 + \epsilon_1 + \epsilon_2)} \\ \times \frac{1}{(a_1 - a_2 + 2\epsilon_1 + \epsilon_2)}, \qquad (2.6)$$

according to the Nekrasov formulae (A.1). In order to get the instanton partition function on $\mathbb{C}^2/\mathbb{Z}_2$ from that on $\mathbb{C}^2$, we must choose the fluctuations that have the appropriate periodicity on $\mathbb{C}^2/\mathbb{Z}_2$. In the case of the trivial holonomy, in order for the fields to be single-valued on $\mathbb{C}^2/\mathbb{Z}_2$, we should choose the fluctuations whose configurations have even parity on $\mathbb{C}^2$. The weights corresponding to the fluctuations that satisfy this condition are ones that are invariant under (2.2) with $p=2$ modulo $2\pi$. In the example above, $(-a_1+a_2), (2\epsilon_1), (-\epsilon_1+\epsilon_2)$ and $(a_1-a_2+\epsilon_1+\epsilon_2)$ satisfy the condition. Then the contribution from the pair of the Young tableaux $(Y_1, Y_2) = (\{1,1\}, \{0\})$ to the instanton partition function on $\mathbb{C}^2/\mathbb{Z}_2$ is

$$\frac{1}{(-a_1+a_2)(2\epsilon_1)(-\epsilon_1+\epsilon_2)(a_1-a_2+\epsilon_1+\epsilon_2)}. \tag{2.7}$$

For a nontrivial holonomy $U$, the periodicity condition is twisted. Then the weights corresponding to the fluctuations with the appropriate periodicity are those invariant modulo $2\pi$ under the simultaneous transformation of (2.2) and (2.3). For example, for the holonomy $U = (1, -1)$, or equivalently $(q_1, q_2) = (0, 1)$, we must extract the weights invariant modulo $2\pi$ under

$$\begin{aligned} a_1 &\to a_1 \\ a_2 &\to a_2 + \pi \\ \epsilon_1 &\to \epsilon_1 + \pi \\ \epsilon_2 &\to \epsilon_2 - \pi, \end{aligned} \tag{2.8}$$

from (2.6). Therefore the contribution from the pair of the Young tableaux $(Y_1, Y_2) = (\{1,1\}, \{0\})$ to the instanton partition function on $\mathbb{C}^2/\mathbb{Z}_2$ is

$$\frac{1}{(-a_1+a_2-\epsilon_1)(2\epsilon_1)(-\epsilon_1+\epsilon_2)(a_1-a_2+2\epsilon_1+\epsilon_2)}. \tag{2.9}$$

To summarize there are two rules in calculating the instanton partition functions with the holonomy $U = (e^{\pi i q_1}, \cdots, e^{\pi i q_N})$ of $U(N)$ gauge theories on $\mathbb{C}^2/\mathbb{Z}_p$. One is which $N$-tuple of Young tableaux we should choose. We should assign charge $q_m + (i-1) - (j-1)$ to the box in th $i$-th column and the $j$-the row in a Young tableaux $Y_m$ ($1 \leq m \leq N$) and choose the $N$-tuples of the Young tableaux that satisfy the relation (2.4), where $k_q$ denotes the number of the boxes with charge $q$ and $N_q$ denotes the number of the diagonal elements with the value $e^{\pi i q}$ in the holonomy $U$. The other rule is which weights we should extract in the contribution from each $N$-tuple of Young tableaux in (A.1). The weights that are invariant modulo $2\pi$ under the transformations (2.2) and (2.3) contribute to the instanton partition functions on $\mathbb{C}^2/\mathbb{Z}_p$.

In the next subsection, we will apply the rules to calculate the instanton partition functions of the $U(2)$ gauge theories on $\mathbb{C}^2/\mathbb{Z}_2$ with and without a fundamental hypermultiplet for the holonomy $U = (1, -1)$ and $(-1, 1)$.

## 2.2 Instanton contributions corresponding to the Ramond sector

Now we calculate the instanton partition function of the sector with the holonomy $U = (1, -1)$ in the $\mathcal{N} = 2$ $U(2)$ pure gauge theory on $\mathbb{C}^2/\mathbb{Z}_2$. As explained below the relation (2.4), this relation becomes $1 = N_1 - 2(k_1 - k_0)$ with $N_1 = 1$, that is, $k_0 = k_1$ in this case and we must choose the pairs of the Young tableaux that satisfy $k_0 = k_1$. Since the instanton number is $k = k_0 + k_1$, we must consider only an even number of instantons. While the 1-instanton factor $\exp(-S_{k=1}^{\text{inst}})$ in the $U(2)$ pure gauge theory on $\mathbb{C}^2$ is $\Lambda^4$ as written in appendix A, that in the gauge theory on $\mathbb{C}^2/\mathbb{Z}_2$ is $\Lambda^2$ since the volume of $\mathbb{C}^2/\mathbb{Z}_2$ is half that of $\mathbb{C}^2$. Therefore the instanton partition function in this case can be written as follows

$$Z_{0,1}^{\text{pure inst}} = \sum_{N \in \mathbb{Z}_{\geq 0}} \Lambda^{4N} Z_{0,1}^{\text{pure},\,(2N)}, \tag{2.10}$$

where $Z_{0,1}^{\text{pure},\,(2N)}$ is the contribution from $2N$ instantons with the holonomy labeled by $(q_1, q_2) = (0, 1)$. First we calculate the 2-instantons partition function. In this case all the pairs of the Young tableaux $(Y_1, Y_2)$ that have two boxes

$$(\{1,1\}, \{0\}),\ (\{2\}, \{0\}),\ (\{0\}, \{1,1\}),\ (\{0\}, \{2\}),$$
$$(\{1\}, \{1\})$$

satisfy the condition (2.4).

Contribution from ({1,1},{0})

Substituting $a_1 = -a_2 = a$ into (2.9), the contribution from $(\{1,1\}, \{0\})$ to the partition function is

$$z_{0,1}^{(\{1,1\},\{0\})} = \frac{1}{(-2a - \epsilon_1)(2\epsilon_1)(-\epsilon_1 + \epsilon_2)(2a + 2\epsilon_1 + \epsilon_2)}, \tag{2.11}$$

where $z_{q_1,q_2}^{(Y_1,Y_2)}$ denotes the contribution from the pair of the Young tableaux $(Y_1, Y_2)$ when the holonomy is $U = (e^{\pi i q_1}, e^{\pi i q_2})$. The contributions $z^{(\{2\},\{0\})}$, $z^{(\{0\},\{1,1\})}$ and $z^{(\{0\},\{2\})}$ can be obtained from (2.11) through the transformations $(\epsilon_1 \leftrightarrow \epsilon_2), (a \leftrightarrow -a)$ and $(\epsilon_1 \leftrightarrow \epsilon_2,\ a \leftrightarrow -a)$. The transformation $(a \leftrightarrow -a)$, or equivalently $(a_1 \leftrightarrow a_2)$ not only changes the pair of the Young tableaux $(Y_1, Y_2) = (\{1,1\}, \{0\})$ to the pair $(Y_1, Y_2) = (\{0\}, \{1,1\})$ but also changes the $\mathbb{Z}_2$ transformation of $a_1$ and $a_2$. Since we would like to fix the holonomy now, the change of the $\mathbb{Z}_2$ transformation may give different results from what we want in a general case. However, in the case of the pure gauge theory, all the fields are in the adjoint representation and the periodicities of them are the same whether $U = (1, -1)$ or $U = (-1, 1)$. Therefore even if we change which of $a_1$ and $a_2$ shifts by $\pi$ under the $\mathbb{Z}_2$ transformation, the partition function does not change.

Contribution from ({1},{1})

Similarly extracting weights invariant modulo $2\pi$ under the $\mathbb{Z}_2$ transformation (2.8) from the summand corresponding to $(\{1\}, \{1\})$ in the Nekrasov formulae (A.1) and substituting

$a_1 = -a_2 = a$ into them

$$(\epsilon_1)^2(a_1 - a_2 + \epsilon_1)(a_1 - a_2 - \epsilon_1)(\epsilon_2)^2(a_1 - a_2 + \epsilon_2)(a_1 - a_2 - \epsilon_2)$$
$$\to (2a + \epsilon_1)(2a - \epsilon_1)(2a + \epsilon_2)(2a - \epsilon_2),$$

we can get the contribution from the pair of the Young tableaux as follows

$$z_{0,1}^{(\{1\},\{1\})} = \frac{1}{(2a+\epsilon_1)(2a-\epsilon_1)(2a+\epsilon_2)(2a-\epsilon_2)}. \tag{2.12}$$

Then the total contribution from 2 instantons for the holonomy $U = (1, -1)$ is

$$\begin{aligned} Z_{0,1}^{\text{pure},(2)} &= z_{0,1}^{(\{1,1\},\{0\})} + (\epsilon_1 \leftrightarrow \epsilon_2) + (a \leftrightarrow -a) + (\epsilon_1 \leftrightarrow \epsilon_2, a \leftrightarrow -a) \\ &\quad + z_{0,1}^{(\{1\},\{1\})} \\ &= -\frac{2(2a^2 - 2\epsilon_1^2 - 5\epsilon_1\epsilon_2 - 2\epsilon_2^2)}{\epsilon_1\epsilon_2(-2a + 2\epsilon_1 + \epsilon_2)(2a + 2\epsilon_1 + \epsilon_2)(-2a + \epsilon_1 + 2\epsilon_2)(2a + \epsilon_1 + 2\epsilon_2)}. \end{aligned} \tag{2.13}$$

Next we calculate the contribution from 4 instantons. The pairs of the Young tableaux that satisfy $k_0 = k_1$ are as follows

$$\begin{aligned} &(\{1,1,1,1\},\{0\}), \quad (\{4\},\{0\}), \quad (\{0\},\{1,1,1,1\}), \quad (\{0\},\{4\}), \\ &(\{2,1,1\},\{0\}), \quad (\{3,1\},\{0\}), \quad (\{0\},\{2,1,1\}), \quad (\{0\},\{3,1\}), \\ &(\{2,2\},\{0\}), \quad\quad\quad\quad\quad\quad\quad\quad (\{0\},\{2,2\}), \\ &(\{1,1,1\},\{1\}), \quad (\{3\},\{1\}), \quad (\{1\},\{1,1,1\}), \quad (\{1\},\{3\}), \\ &(\{1,1\},\{1,1\}), \quad (\{2\},\{2\}), \\ &(\{1,1\},\{2\}, \quad\quad (\{2\},\{1,1\}). \end{aligned} \tag{2.14}$$

Then we can calculate the contributions from these pairs of the Young tableaux in the same way as the case of 2 instantons. They are written in appendix C.1. Summing them up, we get the total contribution from 4 instantons as follows

$$\begin{aligned} Z_{0,1}^{\text{pure},(4)} &= \frac{2\left(16\epsilon_1^4 + 108\epsilon_1^3\epsilon_2 + 202\epsilon_1^2\epsilon_2^2 + 108\epsilon_1\epsilon_2^3 + 16\epsilon_2^4 - a^2(20\epsilon_1^2 + 33\epsilon_1\epsilon_2 + 20\epsilon_2^2) + 4a^4\right)}{\epsilon_1^2\epsilon_2^2(-2a + 2\epsilon_1 + \epsilon_2)(2a + 2\epsilon_1 + \epsilon_2)(-2a + 4\epsilon_1 + \epsilon_2)(2a + 4\epsilon_1 + \epsilon_2)} \\ &\quad \times \frac{1}{(-2a + \epsilon_1 + 2\epsilon_2)(2a + \epsilon_1 + 2\epsilon_2)(-2a + \epsilon_1 + 4\epsilon_2)(2a + \epsilon_1 + 4\epsilon_2)}. \end{aligned} \tag{2.15}$$

We will compare these results (2.13) and (2.15) with super Liouville theory in the next section.

In the case of the holonomy $U = (-1, 1)$, we have to change the $\mathbb{Z}_2$ transformation (2.8). As written below (2.11), in the pure gauge theory the value of the partition function is the same whether $U = (1, -1)$ or $U = (-1, 1)$, that is, $Z_{0,1}^{\text{pure},(k)} = Z_{1,0}^{\text{pure},(k)}$.

If we calculate the instanton partition functions of theories with fundamental hypermultiplets, we can see the difference between the values for the holonomies $U = (1, -1)$ and $U = (-1, 1)$. For example, in the case of the theory with $N_f = 1$ fundamental hypermultiplet, the instanton partition functions with these holonomies can be written as follows

$$Z^{N_f=1}_{q_1,q_2;r} = \sum_{N \in \mathbb{Z}_{\geq 0}} \Lambda^{3N} \, Z^{N_f=1,\,(2N)}_{q_1,q_2;r}, \tag{2.16}$$

where $q_1$ and $q_2$ label the holonomy of the gauge field, $U = (e^{\pi i q_1}, e^{\pi i q_2})$, and $r$ labels the Wilson line for the flavor symmetry around the noncontractible cycle. In the case where $r = 0$ or $1$, we keep invariant weights up to $2\pi$ under the shifts (2.8) together with the shift of the mass of the hypermultiplet, $m \to m + r\pi$. This shift corresponds to the fact that the hypermultiplet fields receive the flavor symmetry transformation when we rotate the fields around the cycle. Using the Nekrasov formulae in appendix A, we can calculate the 2-instanton partition functions as follows

$$Z^{N_f=1,\,(2)}_{0,1;0} = Z^{N_f=1,\,(2)}_{1,0;1}$$
$$= \frac{2(\epsilon_1^3 + \epsilon_2^3) + 7\epsilon_1\epsilon_2(\epsilon_1 + \epsilon_2) - m(4\epsilon_1^2 + 10\epsilon_1\epsilon_2 + 4\epsilon_2^2) + 3a\epsilon_1\epsilon_2 - 2a^2(\epsilon_1 + \epsilon_2 - 2m)}{\epsilon_1\epsilon_2(-2a + 2\epsilon_1 + \epsilon_2)(2a + 2\epsilon_1 + \epsilon_2)(-2a + \epsilon_1 + 2\epsilon_2)(2a + \epsilon_1 + 2\epsilon_2)}$$
(2.17)

$$Z^{N_f=1,\,(2)}_{1,0;0} = Z^{N_f=1,\,(2)}_{0,1;1}$$
$$= \frac{2(\epsilon_1^3 + \epsilon_2^3) + 7\epsilon_1\epsilon_2(\epsilon_1 + \epsilon_2) - m(4\epsilon_1^2 + 10\epsilon_1\epsilon_2 + 4\epsilon_2^2) - 3a\epsilon_1\epsilon_2 - 2a^2(\epsilon_1 + \epsilon_2 - 2m)}{\epsilon_1\epsilon_2(-2a + 2\epsilon_1 + \epsilon_2)(2a + 2\epsilon_1 + \epsilon_2)(-2a + \epsilon_1 + 2\epsilon_2)(2a + \epsilon_1 + 2\epsilon_2)}.$$
(2.18)

The 4-instanton partition functions are given in (C.2). If we change the holonomy of the gauge field as $U \to -U$ and the flavor Wilson line, the periodicities of all the fields including the fundamental fields are the same. Therefore the relations $Z^{N_f=1}_{0,1;0} = Z^{N_f=1}_{1,0;1}$ and $Z^{N_f=1}_{1,0;0} = Z^{N_f=1}_{0,1;1}$ hold.

## 3 Whittaker vectors in the Ramond sector

In a general conformal field theory, eigenstates of the Virasoro generators $L_1$ and $L_2$ in a Verma module are called Whittaker vectors. Now we will consider the Ramond sector in $\mathcal{N} = 1$ super Liouville theory corresponding to $\mathcal{N} = 2$ $U(2)$ gauge theories with the non-$SU(2)$ holonomies on $\mathbb{C}^2/\mathbb{Z}_2$. The Lagrangian of $\mathcal{N} = 1$ super Liouville theory, the super-Virasoro algebra and Ramond primary states are written in appendix B. Then we consider the Verma module for the primary states $|\alpha\rangle_{R\pm}$ in the Ramond sector where $\alpha = a + Q/2$. We propose that the norms of the following Whittaker vectors $|\Lambda^2\rangle_{R\pm}$ coincide with the instanton partition functions with the scalar vev $(a, -a)$ and the non-$SU(2)$ holonomies of the $U(2)$ pure gauge theory.

$$L_1|\Lambda^2\rangle_{R\pm} = \frac{\Lambda^2}{2}|\Lambda^2\rangle_{R\pm} \tag{3.1}$$

$$G_1|\Lambda^2\rangle_{R\pm} = 0. \tag{3.2}$$

Since the second condition implies that $L_2|\Lambda^2\rangle_{R\pm} = 0$, these states are Whittaker vectors. The vectors $|\Lambda^2\rangle_{R\pm}$ can be written as a following power series in $\Lambda^2$

$$|\Lambda^2\rangle_{R\pm} = \sum_{N \in \mathbb{Z}_{\geq 0}} \Lambda^{2N} |N\rangle_{R\pm}, \tag{3.3}$$

where the states $|N\rangle_{R\pm}$ have a level $N$ in the Verma module and satisfy the following conditions

$$L_1|N\rangle_{R\pm} = \frac{1}{2}|N-1\rangle_{R\pm} \tag{3.4}$$

$$G_1|N\rangle_{R\pm} = 0. \tag{3.5}$$

The states $|0\rangle_{R\pm}$, $|1\rangle_{R\pm}$ and $|2\rangle_{R\pm}$ in the Whittaker vectors (3.3) can be written as follows

$$|0\rangle_{R\pm} \equiv |\alpha\rangle_{R\pm} \tag{3.6}$$

$$|1\rangle_{R\pm} = x_1^{(1,\pm)} L_{-1}|0\rangle_{R\pm} + x_2^{(1,\pm)} G_{-1}|0\rangle_{R\mp} \tag{3.7}$$

$$|2\rangle_{R\pm} = x_1^{(2,\pm)} L_{-1}^2|0\rangle_{R\pm} + x_2^{(2,\pm)} L_{-2}|0\rangle_{R\pm}$$
$$+ x_3^{(2,\pm)} G_{-2}|0\rangle_{R\mp} + x_4^{(2,\pm)} L_{-1}G_{-1}|0\rangle_{R\mp}. \tag{3.8}$$

The coefficients $x_1^{(1,\pm)}$ and $x_2^{(1,\pm)}$ in (3.7) are determined by the conditions (3.4) and (3.5) as follows

$$x_1^{(1,\pm)} = \frac{3c + 16\Delta}{4\,(3c\Delta + 16\Delta^2 + 9\beta^2)} \tag{3.9}$$

$$x_2^{(1,\pm)} = \frac{\mp(3 \pm 3i)\beta}{\sqrt{2}\,(3c\Delta + 16\Delta^2 + 9\beta^2)}, \tag{3.10}$$

where $\Delta$ denotes the conformal dimension of the Ramond primary states $\Delta_\alpha^{R\pm}$ in (B.7), $c$ denotes the central charge (B.2) and $\beta = -a/\sqrt{2}$. The coefficients in (3.8) are determined as (C.3). Then if we use BPZ conjugates to determine the norms and choose the convention where $_{R+}\langle\alpha|\alpha\rangle_{R+} = (-i) \times {}_{R-}\langle\alpha|\alpha\rangle_{R-} = 1$, the norms of the states $|1\rangle_{R\pm}$ and $|2\rangle_{R\pm}$ are as follows

$$_{R+}\langle 1|1\rangle_{R+} = (-i) \times {}_{R-}\langle 1|1\rangle_{R-} = \frac{9c^2\Delta + 96c\Delta^2 + 256\Delta^3 + 54c\beta^2 - 288\Delta\beta^2 - 432\beta^4}{8(3c\Delta + 16\Delta^2 + 9\beta^2)^2}. \tag{3.11}$$

and as (C.4). Comparing these norms with the instanton partition functions $Z_{0,1}^{\text{pure},\,(2)}$ (2.13) and $Z_{0,1}^{\text{pure},\,(4)}$ (2.15) and using the fact that $Z_{0,1}^{\text{pure},\,(N)} = Z_{1,0}^{\text{pure},\,(N)}$, we find the following relation

$$Z_{0,1}^{\text{pure},\,(2N)} = Z_{1,0}^{\text{pure},\,(2N)} = {}_{\text{R}+}\langle N|N\rangle_{\text{R}+} = (-i) \times {}_{\text{R}-}\langle N|N\rangle_{\text{R}-}. \tag{3.12}$$

This relation is equivalent to (1.3) because of (2.10) and (3.3).

Next we introduce another set of Whittaker vectors $|\Lambda, m\rangle_{\text{R}\pm}^{(s)}$ and show that the inner products of them and the vectors (3.3) equal to the instanton partition functions of the $U(2)$ gauge theory with $N_f = 1$. The Whittaker vectors $|\Lambda, m\rangle_{\text{R}\pm}^{(s)}$ ($s = 1, 2$) are defined in terms of $c_\pm$ as follows

$$|\Lambda, m\rangle_{\text{R}\pm}^{(s)} = \sum_{N \in \mathbb{Z}_{\geq 0}} \Lambda^N |N, m\rangle_{\text{R}\pm}^{(s)}, \tag{3.13}$$

where the states $|N, m\rangle_{\text{R}\pm}^{(s)}$ are level-$N$ states in the same Verma module as the states $|N\rangle_{\text{R}\pm}$ in (3.3) and satisfy

$$L_1 |N, m\rangle_{\text{R}\pm}^{(s)} = -\left(m - \frac{Q}{2}\right) |N - 1, m\rangle_{\text{R}\pm}^{(s)} \tag{3.14}$$

$$G_1 |N, m\rangle_{\text{R}\pm}^{(s)} = c_\pm |N - 1, m\rangle_{\text{R}\mp}^{(s)}, \tag{3.15}$$

for $N \geq 1$ and $|0, m\rangle_{\text{R}\pm}^{(s)} \equiv |0\rangle_{\text{R}\pm}$. The superscript $(s)$ labels what values we choose as $c_\pm$. We denote the states where we set $c_+ = (1+i)/2$ and $c_- = -(1-i)/2$ by $|\Lambda, m\rangle_{\text{R}\pm}^{(1)}$ and the states where we set $c_+ = -(1+i)/2$ and $c_- = (1-i)/2$ by $|\Lambda, m\rangle_{\text{R}\pm}^{(2)}$. The parameter $m$ is the mass of the fundamental hypermultiplet in the corresponding gauge theory and transformed into $Q - m$ by Weyl symmetry[3].

The states $|1, m\rangle_{\text{R}\pm}^{(s)}$ and $|2, m\rangle_{\text{R}\pm}^{(s)}$ are determined to be as follows

$$|1, m\rangle_{\text{R}\pm}^{(s)} = y_1^{(1,\pm)} L_{-1}|0\rangle_{\text{R}\pm} + y_2^{(1,\pm)} G_{-1}|0\rangle_{\text{R}\mp} \tag{3.16}$$

$$\begin{aligned}|2, m\rangle_{\text{R}\pm}^{(s)} &= y_1^{(2,\pm)} L_{-1}^2 |0\rangle_{\text{R}\pm} + y_2^{(2,\pm)} L_{-2}|0\rangle_{\text{R}\pm} \\ &\quad + y_3^{(2,\pm)} G_{-2}|0\rangle_{\text{R}\mp} + y_4^{(2,\pm)} L_{-1} G_{-1}|0\rangle_{\text{R}\mp},\end{aligned} \tag{3.17}$$

where

$$y_1^{(1,\pm)} = -\frac{(3c + 16\Delta)(m - Q/2) \mp 6(1 \mp i)\sqrt{2} c_\pm \beta}{2(3c\,\Delta + 16\Delta^2 + 9\beta^2)} \tag{3.18}$$

$$y_2^{(1,\pm)} = \frac{8\Delta\, c_\pm \pm 3(1 \pm i)\sqrt{2}\,(m - Q/2)\beta}{3c\,\Delta + 16\Delta^2 + 9\beta^2} \tag{3.19}$$

---

[3]In [3], $m$ is defined to be the mass parameter that is transformed into $-m$ by Weyl symmetry.

and the coefficients in (3.17) are given in (C.5), (C.6), (C.7) and (C.8). Substituting the values of $c_+$ and $c_-$ labeled by $s = 1, 2$, we get the following inner products

$$_{\text{R}+}\langle 1|1, m\rangle^{(1)}_{\text{R}+} = (-i) \times {}_{\text{R}-}\langle 1|1, m\rangle^{(1)}_{\text{R}-} = -\frac{3c(m - Q/2) + 16\Delta(m - Q/2) - 6\sqrt{2}\beta}{4(3c\Delta + 16\Delta^2 + 9\beta^2)} \quad (3.20)$$

$$_{\text{R}+}\langle 1|1, m\rangle^{(2)}_{\text{R}+} = (-i) \times {}_{\text{R}-}\langle 1|1, m\rangle^{(2)}_{\text{R}-} = -\frac{3c(m - Q/2) + 16\Delta(m - Q/2) + 6\sqrt{2}\beta}{4(3c\Delta + 16\Delta^2 + 9\beta^2)} \quad (3.21)$$

and (C.9). Comparing (2.17), (2.18) and (C.2) with (3.20), (3.21) and (C.9), we find the following relations

$$Z^{N_f=1,\,(2N)}_{0,1;\,1} = Z^{N_f=1,\,(2N)}_{1,0;\,0} = \frac{1}{2^N} {}_{\text{R}+}\langle N|N, m\rangle^{(1)}_{\text{R}+} = -\frac{i}{2^N} \times {}_{\text{R}-}\langle N|N, m\rangle^{(1)}_{\text{R}-} \quad (3.22)$$

$$Z^{N_f=1,\,(2N)}_{0,1;\,0} = Z^{N_f=1,\,(2N)}_{1,0;\,1} = \frac{1}{2^N} {}_{\text{R}+}\langle N|N, m\rangle^{(2)}_{\text{R}+} = -\frac{i}{2^N} \times {}_{\text{R}-}\langle N|N, m\rangle^{(2)}_{\text{R}-}. \quad (3.23)$$

These relations are equivalent to (1.4) and (1.5) because of (2.16), (3.3) and (3.13).

## Acknowledgements

The author would like to thank K. Maruyoshi, T. Okuda and M. Taki for helpful discussions. He also thanks T. Okuda for a careful reading of the draft and is also grateful to A. Belavin for reading the manuscript before posting it on the arXiv. This research is supported in part by a JSPS Research Fellowship for Young Scientists.

## A  Nekrasov formulae

We provide the instanton partition function of $\mathcal{N} = 2$ $U(N)$ gauge theory with $N_f$ ($<2N$) fundamental hypermultiplets on $\mathbb{C}^2$. We label a $N$-tuple of Young tableaux by $\vec{Y} = (Y_1, \cdots, Y_N)$. The instanton number is given by the total number of boxes in the $N$-tuple of Young tableaux $|\vec{Y}| = |Y_1| + \cdots + |Y_N|$. The expectation value of the scalar field in the vector multiplet is labeled by the set of the diagonal elements $\vec{a} = (a_1, \cdots, a_N)$ and the mass of the $i$-the fundamental hypermultiplet is denoted by $m_i$, which is transformed into $Q - m$ by Weyl symmetry. The instanton partition function is written as follows

$$Z_{\text{inst}} = \sum_{\vec{Y}} q^{|\vec{Y}|} \frac{\prod_{i=1}^{N_f} z_{\text{fund},i}(\vec{a}, \vec{Y}, m_i)}{z_{\text{vector}}(\vec{a}, \vec{Y})}, \quad (A.1)$$

where $q$ is the 1-instanton factor $\Lambda^{2N-N_f}$ and $\Lambda$ denotes the QCD scale. The partition function has the contributions from the vector multiplet and the $N_f$ fundamental hypermultiplets and they are denoted by $z_{\text{vector}}$ and $z_{\text{fund},i}$ ($1 \leq i \leq N_f$) respectively.

Let $Y_m = \{\lambda_{m,1}, \lambda_{m,2}, \cdots\}$ ($1 \leq m \leq N$) be a Young tableau where $\lambda_{m,i}$ is the height of the $i$-the column. We set $\lambda_{m,i} = 0$ when $i$ is larger than the width of the tableau $Y_m$. Let $Y_m^T = \{\lambda'_{m,1}, \lambda'_{m,2}, \cdots\}$ be its transpose. For a box $s$ in the $i$-the column and the $j$-th row, we define its arm-length $A_{Y_m}(s)$ and leg-length $L_{Y_m}(s)$ with respect to the tableau $Y_m$ by

$$A_{Y_m}(s) = \lambda_{m,i} - j, \quad L_{Y_m}(s) = \lambda'_{m,j} - i. \tag{A.2}$$

Using them, we define a function $E$ by

$$E(a, Y_m, Y_n, s) = a - \epsilon_1 L_{Y_n}(s) + \epsilon_2 (A_{Y_m}(s) + 1). \tag{A.3}$$

The contribution from the vector multiplet $z_{\text{vector}}$ is written as follows

$$z_{\text{vector}}(\vec{a}, \vec{Y}) = \prod_{m,n=1}^{N} \prod_{s \in Y_m} E(a_m - a_n, Y_m, Y_n, s) \prod_{t \in Y_n} (\epsilon_1 + \epsilon_2 - E(a_n - a_m, Y_n, Y_m, t)). \tag{A.4}$$

Note that $L_{Y_n}(s)$ in $E(a_m - a_n, Y_m, Y_n, s)$ is negative when the box $s$ is inside the tableau $Y_m$ but outside the tableau $Y_n$. The contribution from the $i$-th fundamental hypermultiplet $z_{\text{fund},i}$ is written as follows

$$z_{\text{fund},i}(\vec{a}, \vec{Y}, m_i) = \prod_{n=1}^{N} \prod_{s \in Y_n} (\phi(a_n, s) - m_i + \epsilon_1 + \epsilon_2), \tag{A.5}$$

where

$$\phi(a, s) = a + \epsilon_1 (i - 1) + \epsilon_2 (j - 1) \tag{A.6}$$

for the box $s$ in the $i$-th column and the $j$-th row.

# B  Super Liouville theory

The Lagrangian of $\mathcal{N} = 1$ super Liouville theory is as follows

$$\mathcal{L} = \frac{1}{8\pi}(\partial_a \phi)^2 + \frac{1}{2\pi}(\psi \overline{\partial} \psi + \overline{\psi} \partial \overline{\psi}) + 2i\mu b^2 \overline{\psi} \psi e^{b\phi} + 2\pi b^2 \mu^2 e^{2b\phi}. \tag{B.1}$$

The central charge of this conformal field theory is

$$c = 1 + 2Q^2, \quad Q = b + \frac{1}{b}. \tag{B.2}$$

The algebra is as follows

$$[L_n, L_m] = (n-m)L_{n+m} + \frac{c}{8}(n^3 - n)\delta_{n+m}, \tag{B.3}$$

$$\{G_k, G_l\} = 2L_{k+l} + \frac{c}{2}(k^2 - \frac{1}{4})\delta_{k+l}, \tag{B.4}$$

$$[L_n, G_k] = \left(\frac{1}{2}n - k\right) G_{n+k}. \tag{B.5}$$

In the Neveu-Schwarz sector, $n$ and $m$ take integer values and $k$ and $l$ take half-integer values. In the Ramond sector, all of them take integer values.

In the Neveu-Schwarz sector, the primary fields can be written as $V_\alpha^{\text{NS}} = e^{\alpha\phi}$ where $\alpha = \frac{Q}{2} + a$ ($a \in i\mathbb{R}$). The conformal dimensions of them are

$$\Delta_\alpha^{\text{NS}} = \frac{1}{2}\alpha(Q - \alpha). \tag{B.6}$$

In the Ramond sector, the primary fields are written as $V_\alpha^{\text{R}\pm} = \sigma^\pm e^{\alpha\phi}$ with the spin field $\sigma^\pm$ with the dimension $1/16$. The parameter $\alpha$ can be written as $\alpha = \frac{Q}{2} + a$ ($a \in i\mathbb{R}$) as in the Neveu-Schwarz sector. Both of $V_\alpha^{\text{R}\pm}$ have the conformal dimension

$$\Delta_\alpha^{\text{R}} = \frac{1}{16} + \frac{1}{2}\alpha(Q - \alpha), \tag{B.7}$$

and $G_0$ acts on them as follows

$$G_0\, V_\alpha^{\text{R}\pm} = i\beta\, \exp(\mp i\pi/4)\, V_\alpha^{\text{R}\mp},$$

where $\beta = -a/\sqrt{2}$. In section 3 we consider the Ramond primary states $|\alpha\rangle_{\text{R}\pm}$ corresponding to the Ramond primary fields $V_\alpha^{\text{R}\pm}$.

# C  Explicit calculation

## C.1  Instanton partition function

The contribution to the 4-instanton partition function $Z_{0,1}^{\text{pure},\,(4)}$ (2.15) from each pair of the Young tableaux is as follows

$$z_{0,1}^{(\{1,1,1,1\},\{0\})} = \frac{1}{(-2a - 3\epsilon_1)(-2a - \epsilon_1)(2\epsilon_1)(4\epsilon_1)(-3\epsilon_1 + \epsilon_2)(-\epsilon_1 + \epsilon_2)}$$
$$\times \frac{1}{(2a + 2\epsilon_1 + \epsilon_2)(2a + 4\epsilon_1 + \epsilon_2)}$$

$$z_{0,1}^{(\{2,1,1\},\{0\})} = \frac{1}{(-2a - \epsilon_1)(2\epsilon_1)(-2a - \epsilon_2)(3\epsilon_1 - \epsilon_2)(-\epsilon_1 + \epsilon_2)(2a + 2\epsilon_1 + \epsilon_2)}$$
$$\times \frac{1}{(-2\epsilon_1 + 2\epsilon_2)(2a + \epsilon_1 + 2\epsilon_2)}$$

$$z_{0,1}^{(\{2,2\},\{0\})} = \frac{1}{(-2a - \epsilon_1)(2\epsilon_1)(-2a - \epsilon_2)(\epsilon_1 - \epsilon_2)(2\epsilon_2)(-\epsilon_1 + \epsilon_2)}$$
$$\times \frac{1}{(2a + 2\epsilon_1 + \epsilon_2)(2a + \epsilon_1 + 2\epsilon_2)}$$

$$z_{0,1}^{(\{1,1,1\},\{1\})} = \frac{1}{(-2a - \epsilon_1)(2\epsilon_1)(-2a + \epsilon_1)(2a + 3\epsilon_1)(2a + \epsilon_2)(-2a - 2\epsilon_1 + \epsilon_2)}$$
$$\times \frac{1}{(-\epsilon_1 + \epsilon_2)(2a + 2\epsilon_1 + \epsilon_2)}$$

$$z_{0,1}^{(\{1,1\},\{1,1\})} = \frac{1}{(-2a+\epsilon_1)(2a+\epsilon_1)(2\epsilon_1)^2(-2a+\epsilon_2)(2a+\epsilon_2)(-\epsilon_1+\epsilon_2)^2}$$

$$z_{0,1}^{(\{2\},\{1,1\})} = \frac{1}{(2a+\epsilon_1)(2\epsilon_1)(\epsilon_1-\epsilon_2)(-2a+2\epsilon_1-\epsilon_2)(2\epsilon_2)(-2a+\epsilon_2)}$$
$$\times \frac{1}{(-\epsilon_1+\epsilon_2)(2a-\epsilon_1+2\epsilon_2)}.$$

Therefore,

$$\begin{aligned}
Z_{0,1}^{\text{pure},\,(4)} &= z_{0,1}^{(\{1,1,1,1\},\{0\})} + (\epsilon_1 \leftrightarrow \epsilon_2) + (a \leftrightarrow -a) + (\epsilon_1 \leftrightarrow \epsilon_2, a \leftrightarrow -a) \\
&+ z_{0,1}^{(\{2,1,1\},\{0\})} + (\epsilon_1 \leftrightarrow \epsilon_2) + (a \leftrightarrow -a) + (\epsilon_1 \leftrightarrow \epsilon_2, a \leftrightarrow -a) \\
&+ z_{0,1}^{(\{2,2\},\{0\})} + (a \leftrightarrow -a) \\
&+ z_{0,1}^{(\{1,1,1\},\{1\})} + (\epsilon_1 \leftrightarrow \epsilon_2) + (a \leftrightarrow -a) + (\epsilon_1 \leftrightarrow \epsilon_2, a \leftrightarrow -a) \\
&+ z_{0,1}^{(\{1,1\},\{1,1\})} + (\epsilon_1 \leftrightarrow \epsilon_2) \\
&+ z_{0,1}^{(\{2\},\{1,1\})} + (\epsilon_1 \leftrightarrow \epsilon_2) \\
&= \frac{2\left(16\epsilon_1^4 + 108\epsilon_1^3\epsilon_2 + 202\epsilon_1^2\epsilon_2^2 + 108\epsilon_1\epsilon_2^3 + 16\epsilon_2^4 - a^2(20\epsilon_1^2 - 20\epsilon_2^2 - 33\epsilon_1\epsilon_2) + 4a^4\right)}{\epsilon_1^2\epsilon_2^2(-2a+2\epsilon_1+\epsilon_2)(2a+2\epsilon_1+\epsilon_2)(-2a+4\epsilon_1+\epsilon_2)(2a+4\epsilon_1+\epsilon_2)} \\
&\times \frac{1}{(-2a+\epsilon_1+2\epsilon_2)(2a+\epsilon_1+2\epsilon_2)(-2a+\epsilon_1+4\epsilon_2)(2a+\epsilon_1+4\epsilon_2)}.
\end{aligned} \quad \text{(C.1)}$$

In the case of the $U(2)$ gauge theory with $N_f = 1$, the 4-instanton partition functions with the holonomies of the gauge field $U = (e^{\pi i q_1}, e^{\pi i q_2})$ and the flavor Wilson line labeled by $r$, $Z_{q_1,q_2,;r}^{(4)}$ are as follows

$$\begin{aligned}
Z_{0,1;1}^{(4)} &= Z_{1,0;0}^{(4)} \\
&= \Big(32\epsilon_1^6 + 304\epsilon_1^5\epsilon_2 + 1030\epsilon_1^4\epsilon_2^2 + 1543\epsilon_1^3\epsilon_2^3 + 1030\epsilon_1^2\epsilon_2^4 + 304\epsilon_1\epsilon_2^5 + 32\epsilon_2^6 \\
&\quad + m(-128\epsilon_1^5 - 992\epsilon_1^4\epsilon_2 - 2480\epsilon_1^3\epsilon_2^2 - 2480\epsilon_1^2\epsilon_2^3 - 992\epsilon_1\epsilon_2^4 - 128\epsilon_2^5) \\
&\quad + m^2(128\epsilon_1^4 + 864\epsilon_1^3\epsilon_2 + 1616\epsilon_1^2\epsilon_2^2 + 864\epsilon_1\epsilon_2^3 + 128\epsilon_2^4) \\
&\quad + a\big(-96\epsilon_1^4\epsilon_2 - 294\epsilon_1^3\epsilon_2^2 - 294\epsilon_1^2\epsilon_2^3 - 96\epsilon_1\epsilon_2^4 + m(192\epsilon_1^3\epsilon_2 + 396\epsilon_1^2\epsilon_2^2 + 192\epsilon_1\epsilon_2^3)\big) \\
&\quad + a^2\big(-40\epsilon_1^4 - 248\epsilon_1^3\epsilon_2 - 410\epsilon_1^2\epsilon_2^2 - 248\epsilon_1\epsilon_2^3 - 40\epsilon_2^4 \\
&\qquad + m(160\epsilon_1^3 + 424\epsilon_1^2\epsilon_2 + 424\epsilon_1\epsilon_2^2 + 160\epsilon_2^3) + m^2(-160\epsilon_1^2 - 264\epsilon_1\epsilon_2 - 160\epsilon_2^2)\big) \\
&\quad + a^3\big(24\epsilon_1^2\epsilon_2 + 24\epsilon_1\epsilon_2^2 - 48m\epsilon_1\epsilon_2\big) \\
&\quad + a^4\big(8\epsilon_1^2 + 40\epsilon_1\epsilon_2 + 8\epsilon_2^2 + m(-32\epsilon_1 - 32\epsilon_2) + 32m^2\big)\Big) \\
&\times \Big(4\epsilon_1^2\epsilon_2^2(-2a+2\epsilon_1+\epsilon_2)(2a+2\epsilon_1+\epsilon_2)(-2a+4\epsilon_1+\epsilon_2)(2a+4\epsilon_1+\epsilon_2) \\
&\quad (-2a+\epsilon_1+2\epsilon_2)(2a+\epsilon_1+2\epsilon_2)(-2a+\epsilon_1+4\epsilon_2)(2a+\epsilon_1+4\epsilon_2)\Big)^{-1}. \quad \text{(C.2)}
\end{aligned}$$

If we transform $a$ into $-a$, we get the value of the instanton partition functions $Z_{1,0;1}^{(4)} = Z_{0,1;0}^{(4)}$.

## C.2 Whittaker vectors

We denote $\Delta_\alpha^{\text{R}}$ just by $\Delta$ in this subsection. Note that we choose the convention where the coefficient of the central charge in (B.3) is $1/8$ rather than $1/12$. The coefficients in (3.8) are as follows

$$
\begin{aligned}
x_1^{(2,\pm)} &= \frac{45c^2 + 24c\,(7+12\Delta) + 32\,(28\Delta + 8\Delta^2 + 45\beta^2)}{(3c\Delta + 16\Delta^2 + 9\beta^2)\,(15c(7+32\Delta) + 8\,(49 + 238\Delta + 64\Delta^2 + 450\beta^2))} \\
x_2^{(2,\pm)} &= -\frac{3(168c + 45c^2 + 896\Delta + 288c\Delta + 256\Delta^2 + 2112\beta^2 - 720c\beta^2 - 768\Delta\beta^2)}{8(3c\Delta + 16\Delta^2 + 9\beta^2)(392 + 105c + 1904\Delta + 480c\Delta + 512\Delta^2 + 3600\beta^2)} \\
x_3^{(2,\pm)} &= \frac{\mp(3\pm 3i)\beta\,(-75c + 8(21 + 46\Delta + 90\beta^2))}{\sqrt{2}(3c\Delta + 16\Delta^2 + 9\beta^2)\,(15c\,(7+32\Delta) + 8\,(49 + 238\Delta + 64\Delta^2 + 450\beta^2))} \\
x_4^{(2,\pm)} &= \frac{\mp(12 \pm 12i)\sqrt{2}(-14 + 15c + 16\Delta)\beta}{(3c\Delta + 16\Delta^2 + 9\beta^2)(15c(7+32\Delta) + 8(49 + 238\Delta + 64\Delta^2 + 450\beta^2))}. \quad\text{(C.3)}
\end{aligned}
$$

Then the norms of the states $|2\rangle_{\text{R}\pm}$ (3.8) are as follows

$$
\begin{aligned}
{}_{\text{R}+}\langle 2|2\rangle_{\text{R}+} &= (-i) \times {}_{\text{R}-}\langle 2|2\rangle_{\text{R}-} \\
&= \big(1008c^2\Delta + 270c^3\Delta + 10752c\Delta^2 + 3168c^2\Delta^2 + 28672\Delta^3 + 10752c\Delta^3 + 8192\Delta^4 \\
&\quad + 6048c\beta^2 + 405c^2\beta^2 - 32256\Delta\beta^2 + 31968c\Delta\beta^2 - 135936\Delta^2\beta^2 - 22464\beta^4 \\
&\quad + 25920c\beta^4 - 497664\Delta\beta^4 - 311040\beta^6\big) \\
&\quad \times \big(8(3c\Delta + 16\Delta^2 + 9\beta^2)^2(392 + 105c + 1904\Delta + 480c\Delta + 512\Delta^2 + 3600\beta^2)\big)^{-1}. \quad\text{(C.4)}
\end{aligned}
$$

The coefficients in (3.17) are as follows. We denote $m - Q/2$ by $m_{sL}$.

$$
\begin{aligned}
y_1^{(2,\pm)} &= 4\Big(45c^2 m_{sL}^2 + 128\Delta^2(2m_{sL}^2 - 3c_+c_-) + 64\Delta\big(14m_{sL}^2 - 21c_+c_- \mp 3(1\mp i)\sqrt{2}m_{sL}c_\pm\beta\big) \\
&\quad + 12c\big(2(7+12\Delta)m_{sL}^2 - 30\Delta c_+c_- \mp 15(1\mp i)\sqrt{2}m_{sL}c_\pm\beta\big) \\
&\quad + 24\beta\big(\pm 7\sqrt{2}(1\mp i)m_{sL}c_\pm + 60m_{sL}^2\beta - 75c_+c_-\beta\big)\Big) \\
&\quad \times \Big((3c\Delta + 16\Delta^2 + 9\beta^2)\big(15c(7+32\Delta) + 8(49 + 238\Delta + 64\Delta^2 + 450\beta^2)\big)\Big)^{-1} \quad\text{(C.5)}
\end{aligned}
$$

$$
\begin{aligned}
y_2^{(2,\pm)} &= \Big(c\big(1920\Delta(1+2\Delta)c_+c_- \pm 540(1\mp i)\sqrt{2}m_{sL}c_\pm\beta - 72m_{sL}^2(7+12\Delta - 30\beta^2)\big) \\
&\quad - 135c^2 m_{sL}^2 + 4096\Delta^3 c_+c_- - 256\Delta^2(3m_{sL}^2 - 64c_+c_-) \\
&\quad + 48\beta\big(-132m_{sL}^2\beta + 165c_+c_-\beta \mp (5\mp 5i)\sqrt{2}m_{sL}c_\pm(7+36\beta^2)\big) \\
&\quad + 16\Delta\big(300(\mp 1+i)\sqrt{2}m_{sL}c_\pm\beta + 24m_{sL}^2(-7+6\beta^2) + 32c_+c_-(14+45\beta^2)\big)\Big) \\
&\quad \times \Big(2(3c\Delta + 16\Delta^2 + 9\beta^2)(15c(7+32\Delta) + 8(49 + 238\Delta + 64\Delta^2 + 450\beta^2))\Big)^{-1} \quad\text{(C.6)}
\end{aligned}
$$

$$y_3^{(2,\pm)}$$
$$= 2(1\pm i)\bigg(3c\,m_{sL}\big(4(1\mp i)(7+32\Delta)c_\pm \pm 75\sqrt{2}m_{sL}\beta\big)$$
$$\pm 256\Delta^2 c_\pm\big(4(\mp 1+i)m_{sL} - 3\sqrt{2}c_\mp\beta\big)$$
$$\mp 24\beta\big(-14\sqrt{2}c_+c_- \mp 24(1\mp i)m_{sL}c_\pm\beta + 3\sqrt{2}m_{sL}^2(7+30\beta^2)\big)$$
$$\mp 16\Delta\big(69\sqrt{2}m_{sL}^2\beta + 27\sqrt{2}c_+c_-\beta \pm 2(1\mp i)m_{sL}c_\pm(7+72\beta^2)\big)\bigg)$$
$$\times \Big((3c\Delta + 16\Delta^2 + 9\beta^2)\big(15c(7+32\Delta) + 8(49+238\Delta+64\Delta^2+450\beta^2)\big)\Big)^{-1} \quad \text{(C.7)}$$

$$y_4^{(2,\pm)}$$
$$= -4\bigg(15c\,m_{sL}\big((7+32\Delta)c_\pm \pm 12(1\pm i)\sqrt{2}m_{sL}\beta\big)$$
$$+ 2(1\pm i)\Big(128(1\mp i)\Delta^2 m_{sL}c_\pm + 3\beta\big(\mp 28\sqrt{2}m_{sL}^2 \pm 35\sqrt{2}c_+c_- + 60(1\mp i)m_{sL}c_\pm\beta\big)$$
$$+ \Delta\big(28(1\mp i)m_{sL}c_\pm \pm 96\sqrt{2}m_{sL}^2\beta \mp 240\sqrt{2}c_+c_-\beta\big)\Big)\bigg)$$
$$\times \Big((3c\Delta + 16\Delta^2 + 9\beta^2)\big(15c(7+32\Delta) + 8(49+238\Delta+64\Delta^2+450\beta^2)\big)\Big)^{-1} \quad \text{(C.8)}$$

If we set $c_+ = (1+i)/2$ and $c_- = -(1-i)/2$, the inner products in (3.22) for $N=2$ are as follows

$$_{R+}\langle 2|2,m\rangle_{R+}^{(1)} = (-i) \times {}_{R-}\langle 2|2,m\rangle_{R-}^{(2)}$$
$$= \Big(129024c\Delta^2 + 34560c^2\Delta^2 + 688128\Delta^3 + 221184c\Delta^3 + 196608\Delta^4 + m_{sL}^2(32256c^2\Delta$$
$$+ 8640c^3\Delta + 344064c\Delta^2 + 101376c^2\Delta^2 + 917504\Delta^3 + 344064c\Delta^3 + 262144\Delta^4)$$
$$+ \beta\,m_{sL}(-4536\sqrt{2}c^2 - 1215\sqrt{2}c^3 + 145152\sqrt{2}c\Delta - 31536\sqrt{2}c^2\Delta - 473088\sqrt{2}\Delta^2$$
$$+ 283392\sqrt{2}c\Delta^2 - 4067328\sqrt{2}\Delta^3)$$
$$+ \beta^2\big(18144c + 4860c^2 + 96768\Delta + 127872c\Delta + 25920c^2\Delta + 2165760\Delta^2 - 387072c\Delta^2$$
$$- 442368\Delta^3 + m_{sL}^2(193536c + 12960c^2 - 1032192\Delta + 1022976c\Delta - 4349952\Delta^2)\big)$$
$$+ \beta^3 m_{sL}\big(96768\sqrt{2} - 88128\sqrt{2}c + 38880\sqrt{2}c^2 + 156672\sqrt{2}\Delta + 387072\sqrt{2}c\Delta$$
$$- 10838016\sqrt{2}\Delta^2\big)$$
$$+ \beta^4\big(228096 - 77760c + 1133568\Delta - 414720c\Delta - 442368\Delta^2 + m_{sL}^2(-718848$$
$$+ 829440c - 15925248\Delta)\big)$$
$$+ \beta^5 m_{sL}(912384\sqrt{2} - 311040\sqrt{2}c - 6967296\sqrt{2}\Delta) - 9953280\beta^6 m_{sL}^2\Big)$$
$$\times \big(64(3c\Delta + 16\Delta^2 + 9\beta^2)^2(392 + 105c + 1904\Delta + 480c\Delta + 512\Delta^2 + 3600\beta^2)\big)^{-1}. \quad \text{(C.9)}$$

If we transform $\beta$ into $-\beta$, we get the value of ${}_{\text{R}+}\langle 2|2,m\rangle^{(2)}_{\text{R}+} = (-i) \times {}_{\text{R}-}\langle 2|2,m\rangle^{(2)}_{\text{R}-}$, where we set $c_+ = -(1+i)/2$ and $c_- = (1-i)/2$.